\documentclass[12pt]{article} 
\usepackage{amssymb} 
\makeatletter
\@addtoreset{equation}{section}

\makeatother

\begin{document}
\title{Distributional Energy-Momentum Densities of Schwarzschild 
Space-Time}
\author{Toshiharu \textsc{Kawai}\thanks{%
    Electronic address: \texttt{kawai@sci.osaka-cu.ac.jp}} 
  and Eisaku \textsc{Sakane}\thanks{%
    Electronic address: \texttt{sakane@sci.osaka-cu.ac.jp}}\\
\emph{Department of Physics, Osaka City University, Sugimoto,}\\
\emph{Sumiyoshi-ku, Osaka 558}}
\date{}
\maketitle

\begin{abstract}
  For Schwarzschild space-time, distributional expressions of
  energy-momentum densities and of scalar concomitants of the
  curvature tensors are examined for a class of coordinate systems
  which includes those of the Schwarzschild and of Kerr-Schild types
  as special cases. The energy-momentum density
  $\widetilde{\mbox{\boldmath $T$}}_\mu^{\ \nu}(x)$ of the
  gravitational source and the gravitational energy-momentum
  pseudo-tensor density $\widetilde{\mbox{\boldmath $t$}}_\mu^{\ \nu}$
  have the expressions $\widetilde{\mbox{\boldmath $T$}}_\mu^{\ 
    \nu}(x) =-Mc^2\delta_\mu^{\ 0}\delta_0^{\ \nu}
  \delta^{(3)}(\mbox{\boldmath $x$})$ and $\widetilde{\mbox{\boldmath
      $t$}}_\mu^{\ \nu}=0$, respectively.  In expressions of the
  curvature squares for this class of coordinate systems, there are
  terms like $\delta^{(3)}(\mbox{\boldmath $x$})/r^3$ and
  $\left[\delta^{(3)}(\mbox{\boldmath $x$)}\right]^2$, as well as
  other terms, which are singular at $\mbox{\boldmath
    $x$}=\mbox{\boldmath $0$}$. It is pointed out that the well-known
  expression $R^{\rho\sigma\mu\nu}(\{\}) R_{\rho\sigma\mu\nu}(\{\})$
  \linebreak $=48G^{2}M^{2}/c^{4}r^{6}$ is not correct, if we define
  $1/r^6\stackrel{\mbox{\scriptsize def}}{=}
  \lim_{\epsilon\to0}1/(r^2+\epsilon^2)^3$.
\end{abstract}
\pagebreak

\section{Introduction}
\label{sec:1}

\hspace*{3ex} In general relativity, many investigations have been
made with regard to exact solutions of the Einstein equation and the
singularity structure of space-time, but a distribution theoretical
treatment of these has not been developed sufficiently.  This is the
case even for the well-known Schwarzschild solution, which is given
by, in the Schwarzschild coordinates $(\hat{x}^{0}, \hat{r}, \theta,
\phi)$,
\begin{equation}
  \label{eq:Schwarzschild-metric}
  ds^{2}=-\left(1-\frac{a}{\hat{r}}\right)(d\hat{x}^{0})^{2}
  +\left(1-\frac{a}{\hat{r}}\right)^{-1}(d\hat{r})^2
  +\hat{r}^2\left[(d\theta )^2
  +{\sin^{2}\theta}(d\phi )^2\right].
\end{equation}
Here, $a$ is the Schwarzschild radius $a=2GM/c^{2}$ with $G, M$ and
$c$ being the Newton gravitational constant, mass of the source, and
the light velocity in vacuum Minkowski space-time, respectively. This
metric, which is usually called the exterior Schwarzschild metric, is
a solution of the Einstein equation in the vacuum region outside a
static, spherically symmetric, extended gravitating body, and it is
meaningless inside the body.

The metric (\ref{eq:Schwarzschild-metric}) can also describe the
gravitational field produced by a point-like particle located at
$\hat{\mbox{\boldmath $x$}}=\mbox{\boldmath $0$}$. When we say, on the
basis of the expression of the curvature square
$R^{\rho\sigma\mu\nu}(\{\})R_{\rho\sigma\mu\nu}(\{\})$ obtained from
the metric (\ref{eq:Schwarzschild-metric}), that $\hat{\mbox{\boldmath
    $x$}}=\mbox{\boldmath $0$}$ is a singularity of the Schwarzschild
space-time, the source is considered to be point-like and this metric
is regarded as meaningful everywhere in space-time.  Then, we have
\begin{eqnarray}
  \hat{\widetilde{\mbox{\boldmath $T$}}}_\mu^{\raisebox{-.6ex}%
    {\scriptsize $\;\,\nu$}}(\hat{x})=0\;,
  \qquad \mbox{for}\ \hat{\mbox{\boldmath $x$}}\neq\mbox{\boldmath $0$}\;,
  \nonumber\\
  \hat{\widetilde{\mbox{\boldmath $t$}}}_{\mu}^{\raisebox{-.6ex}%
    {\scriptsize $\;\,\nu$}}(\hat{x})=0\;,
  \qquad \hat{P}_{\mu}=(-Mc^{2}, 0, 0, 0)\;,
\end{eqnarray}
where $\hat{\widetilde{\mbox{\boldmath $T$}}}_\mu^{\raisebox{-.6ex}%
  {\scriptsize $\;\,\nu$}}$ and
$\hat{\widetilde{\mbox{\boldmath $t$}}}_{\mu}^{\raisebox{-.6ex}%
  {\scriptsize $\;\,\nu$}}$ represent the energy-momentum densities
of the gravity source and of the gravitational field, respectively,
and $\hat{P}_{\mu}$ is the total energy-momentum of the system.
Thus, it is unambiguously expected that the 
energy-momentum density 
$\hat{\widetilde{\mbox{\boldmath $T$}}}_\mu^{\raisebox{-.6ex}%
  {\scriptsize $\;\,\nu$}}$ of the gravity source has the expression
\begin{equation}
  \label{eq:energy-momentum-density}
  \hat{\widetilde{\mbox{\boldmath $T$}}}_\mu^{\raisebox{-.6ex}%
    {\scriptsize $\;\,\nu$}}(\hat{x})
  =-Mc^2\delta_\mu^{\ 0}\delta_0^{\ \nu}
  \delta^{(3)}(\hat{\mbox{\boldmath $x$}})\;.
\end{equation}
Nevertheless, this has not been given explicitly in any textbook of
general relativity, as far as the present authors know.  Also, in the
literature, \emph{without giving distribution theoretical
  examinations}, the expression
$R^{\rho\sigma\mu\nu}(\{\})R_{\rho\sigma\mu\nu}(\{\}) =12a^{2}/r^{6}$
is given, and the scalars $R^{\mu\nu}(\{\})R_{\mu\nu}(\{\})$ and
$R(\{\})$ are not given explicitly, where ${R^{\rho }}_{\mu \nu
  \lambda }(\{\})$, $R_{\mu \nu }(\{\})$ and $R(\{\})$ stand for the
Riemann-Christoffel curvature tensor, Ricci tensor and scalar
curvature, respectively.

Recently, distributional expressions of energy-momentum densities of
gravitational \linebreak sources for the Schwarzschild and
Kerr-Newmann space-times have been given by Balasin and
Nachbagauer\cite{Balasin1,Balasin2,Balasin3,Balasin4} for the
Kerr-Schild coordinate system.
 
The purpose of this paper is to examine, from the distribution
theoretical point of view, the energy-momentum densities, and the
scalar concomitants 
$R^{\rho\sigma\mu\nu}(\{\})\times R_{\rho\sigma\mu\nu}(\{\})$, 
$R^{\mu\nu}(\{\})R_{\mu\nu}(\{\})$ and $R(\{\})$ of the Schwarzschild
space-time by assuming that the metric (\ref{eq:Schwarzschild-metric})
is meaningful everywhere in space-time.  The examination is given for
the coordinates $(\hat{x}^{0}, \hat{r}, \theta, \phi )$ and for the
class of the coordinates $(x^{0},r,\theta ,\phi)$ obtained by the
transformation
\begin{equation}
  \label{eq:coordinate-transformation}
  \hat{x}^{0}=\Omega x^{0}+f(r)\;, \; \; \hat{r}=K(r)
\end{equation}
with $\Omega $ being a constant and $f$ and $K$ being functions
of the coordinate $r$. (Note the following: (a)We obtain Kerr-Schild 
coordinates when $f(r)=a\ln |a-r|, K(r)=r$ and $\Omega =1$. (b)When 
$f'(r)=0, K(r)=r+a/2$ and $\Omega =1$, we obtain Lanczos coordinates
which satisfy the harmonic condition.) 

\section{Preliminary}
\label{sec:2}

\hspace*{3ex} We briefly summarize the basics of general 
relativity, as a preliminary to latter discussion.

In general relativity, the space-time is assumed to be a
four-dimensional differentiable manifold endowed with the Lorentzian
metric $ds^{2}=g_{\mu \nu }dx^{\mu}dx^{\nu}$ $(\mu,\nu=0,1,2,3)$.  At
each point $p$ of space-time, the metric can be diagonalized as
$ds^{2}_{p}=\eta_{\mu \nu}(dX^{\mu })_{p}(dX^{\nu })_{p}$ with
$(\eta_{\mu \nu })\stackrel{\mbox{\scriptsize def}}{=}
\mbox{diag}\;(-1,1,1,1)$, by choosing the coordinate system
$\{X^{\mu};\mu =0,1,2,3\}$ appropriately.

The curvature tensor is given by 
\begin{equation} 
{R^\rho}_{\sigma\mu\nu}(\{\})
\stackrel{\mbox{\scriptsize def}}{=}
\partial_\mu{\rho\brace\sigma\ \nu}
-\partial_\nu{\rho\brace\sigma\ \mu}
+{\rho\brace\lambda\ \mu}{\lambda\brace\sigma\ \nu}
-{\rho\brace\lambda\ \nu}{\lambda\brace\sigma\ \mu}
\end{equation}
with ${\rho\brace\sigma\ \nu}$ being the Christoffel symbol. The
fundamental action integral ${\Bbb I}$ is 
\begin{equation}
{\Bbb I} 
\stackrel{\mbox{\scriptsize def}}{=}
\frac{1}{c}\int \,(\bar{\Bbb L}_{G}+{\Bbb L}_{M})d^4x\;, \\
\end{equation}
where ${\Bbb L}_{M}$ is the Lagrangian density of a gravitational 
source and $\bar{\Bbb L}_{G}$ is the gravitational Lagrangian density
given by 
\begin{equation}
\bar{\Bbb L}_G 
\stackrel{\mbox{\scriptsize def}}{=} \frac{1}{2\kappa}{\Bbb G}\;.
\end{equation}  
Here $\kappa $ is the Einstein gravitational constant 
$\kappa =8\pi G/c^{4}$ and ${\Bbb G}$ is defined by 
\begin{equation}
{\Bbb G} 
\stackrel{\mbox{\scriptsize def}}{=} 
\sqrt{-g}g^{\mu\nu}\left(%
{\lambda\brace\mu\ \rho}{\rho\brace\nu\ \lambda}
-{\lambda\brace\mu\ \nu}{\rho\brace\lambda\ \rho}\right)
\end{equation}
with $g\stackrel{\mbox{\scriptsize def}}{=}\det (g_{\mu \nu })$.
There exists the relation 
\begin{equation}
\sqrt{-g}R(\{\})={\Bbb G}+\partial_{\mu }{\Bbb D}^{\mu}\;,
\end{equation}
with
\begin{equation}
{\Bbb D}^{\mu }\stackrel{\mbox{\scriptsize def}}{=}-\sqrt{-g}
\left(g^{\mu \nu}{\lambda\brace\nu\ \lambda}
-g^{\nu \lambda}{\mu\brace\nu\ \lambda}\right)\;.
\end{equation}
Also, we have defined the scalar curvature 
by\footnote{Raising and lowering the indices $\mu ,\nu ,
\lambda , \cdots $ is accomplished with the aid of 
$(g^{\mu \nu })\stackrel{\mbox{\scriptsize def}}{=}
(g_{\mu \nu })^{-1}$ and $(g_{\mu \nu })$.}
\begin{equation}
R(\{\})\stackrel{\mbox{\scriptsize def}}{=}{R^{\mu}}_{ \mu }(\{\})
\end{equation}
with the Ricci tensor 
\begin{equation}
R_{\mu\nu}(\{\}) \stackrel{\mbox{\scriptsize def}}{=} 
{R^\lambda}_{\mu\lambda\nu}(\{\})\;.
\end{equation}
From the action ${\Bbb I}$, the Einstein equation
\begin{equation}
{G_{\mu}}^{\nu}(\{\})\stackrel{\mbox{\scriptsize def}}{=} 
{R_{\mu}}^{\nu}(\{\})-\frac{1}{2}{\delta_{\mu}}^{\nu}R(\{\})
=\kappa {T_{\mu }}^{\nu }\;,
\end{equation}
follows, where ${T_{\mu }}^{\nu }$ is defined by
\begin{equation}
{T_{\mu }}^{\nu }\stackrel{\mbox{\scriptsize def}}{=}
\frac{{\widetilde{\mbox{\boldmath $T$}}_{\mu }}^{\ \nu}}{\sqrt{-g}}
\end{equation}
with 
\begin{equation}
\widetilde{\mbox{\boldmath $T$}}_\mu ^{\ \nu }
\stackrel{\mbox{\scriptsize def}}{=}
2g_{\mu \lambda }\frac{\delta {\Bbb L}_{M}}
{\delta g_{\lambda \nu }}
\end{equation}
being the energy-momentum density of the gravity source.  The
energy-momentum pseudo-tensor density\footnote{ One may say, \lq\lq It
  is meaningless to deal with $\widetilde{\mbox{\boldmath $t$}}_\mu^{\ 
    \nu}$, because this is not a physically sensible local object."
  However, the following should be pointed out: (a)For an isolated
  system, $\widetilde{\mbox{\boldmath $t$}}_\mu^{\ \nu}$ for an
  asymptotically Minkowskian coordinate system gives the total
  gravitational energy-momentum when integrated over a space-like
  surface.  (b)Our main concern is not $\widetilde{\mbox{\boldmath
      $t$}}_\mu^{\ \nu}$, and it plays only an auxiliary role in our
  discussion (see also the footnote \ref{footnote:auxiliary} (a) on
  page \pageref{footnote:auxiliary}).}  $\widetilde{\mbox{\boldmath
    $t$}}_\mu^{\ \nu}$ of the gravitational field is defined by
\begin{equation}
\widetilde{\mbox{\boldmath $t$}}_\mu^{\ \nu}
\stackrel{\mbox{\scriptsize def}}{=}
{\delta_\mu}^\nu \bar{\Bbb L}_G
-\frac{\partial\bar{\Bbb L}_G}
{\partial g_{\sigma\tau,\nu}}g_{\sigma\tau,\mu}
\end{equation}
with $g_{\sigma \tau,\nu}\stackrel{\mbox{\scriptsize def}}{=}
\partial g_{\sigma \tau}/\partial x^{\nu }$.

\section{Schwarzschild coordinates}
\label{sec:3}

\hspace*{3ex} Let us first consider the problem in Schwarzschild
coordinates $(\hat{x}^{0},\hat{r},\theta, \phi)$ by which the
Schwarzschild metric is expressed as Eq. (1$\cdot $1).  By using the
Cartesian coordinates $(\hat{x}^{0},
\hat{x}^{1},\hat{x}^{2},\hat{x}^{3})$, which are related to
$(\hat{x}^{0},\hat{r},\theta, \phi)$ through the relation
\begin{equation}
\hat{x}^{1}=\hat{r}\cos \phi \sin \theta \;, \; \;
\hat{x}^{2}=\hat{r}\sin \phi \sin \theta \;,\; \;
\hat{x}^{3}=\hat{r}\cos \theta \;,
\end{equation}
the metric takes the form
\begin{equation}
ds^{2}={\hat{g}}_{\mu \nu }d\hat{x}^{\mu }d\hat{x}^{\nu }\;,
\end{equation}
where $\hat{g}_{\mu \nu }$ is given by
\begin{eqnarray}
{\hat{g}}_{00}&=&-(1-h)\;, \; \; {\hat{g}}_{0 \alpha }=0\;,
\nonumber \\
{\hat{g}}_{\alpha \beta}&=&\delta^{\alpha \beta}
+h(1-h)^{-1}\frac{\hat{x}^{\alpha }\hat{x}^{\beta }}{\hat{r}^{2}}
\;,\;\;\; \alpha,\beta=1,2,3
\end{eqnarray}
with $h \stackrel{\mbox{\scriptsize def}}{=}a/\hat{r}$.

We know that
\begin{eqnarray}
\kappa\hat{\widetilde{\mbox{\boldmath $T$}}}_0^{\raisebox{-.6ex}%
{\scriptsize $\;\,0$}}&=&-\frac{h'}{\hat{r}}-\frac{h}{\hat{r}^2}\;,
\nonumber\\
\kappa \hat{\widetilde{\mbox{\boldmath $T$}}}_0^{\raisebox{-.6ex}%
{\scriptsize $\;\,\alpha$}}&=&0\;, \; \; \; 
\kappa\hat{\widetilde{\mbox{\boldmath $T$}}}_{\alpha}^{%
\raisebox{-.6ex}{\scriptsize $\;\,0$}}=0\;, \nonumber \\
\kappa\hat{\widetilde{\mbox{\boldmath $T$}}}_\alpha^{%
\raisebox{-.6ex}{\scriptsize $\;\,\beta$}}&=&
\delta_\alpha^{\ \beta}\left(-\frac{h''}{2}
-\frac{h'}{\hat{r}}\right)
+\frac{\hat{x}^\alpha \hat{x}^\beta}{\hat{r}^2}
\left(\frac{h''}{2}-\frac{h}{\hat{r}^2}\right)\;,
\end{eqnarray}
\begin{equation}
{\hat{\Bbb G}}=0\;,\; \; \;
\hat{\widetilde{\mbox{\boldmath $t$}}}_\mu^{\raisebox{-.6ex}%
{\scriptsize $\;\,\nu$}}=0\;,
\end{equation}
where the hatted symbols $\hat{\Bbb G},
\hat{\widetilde{\mbox{\boldmath $t$}}}_\mu^{\raisebox{-.6ex}%
{\scriptsize $\;\,\nu$}}$ and
$\hat{\widetilde{\mbox{\boldmath $T$}}}_\mu^{\raisebox{-.6ex}%
{\scriptsize $\;\,\nu$}}$ represent, respectively, ${\Bbb G},
\widetilde{{\mbox{\boldmath $t$}}}_\mu^{\ \nu}$ and
$\widetilde{{\mbox{\boldmath $T$}}}_\mu^{\ \nu }$ in the coordinate
system $\{\hat{x}^{\mu };\mu=0,1,2,3\}$. Also, we have defined 
$h'\stackrel{\mbox{\scriptsize def}}{=}dh/d\hat{r}$ and 
$h''\stackrel{\mbox{\scriptsize def}}{=}d^{2}h/d\hat{r}^{2}$.

Regularizing the function $h=a/\hat{r}$ as 
$a/\sqrt{\hat{r}^{2}+\epsilon^{2}}$ with $\epsilon $ being a 
real number, we obtain
\begin{eqnarray}
\kappa\hat{\widetilde{\mbox{\boldmath $T$}}}_0^{\raisebox{-.6ex}%
{\scriptsize $\;\,0$}}(\hat{x};\epsilon)&=&
-\frac{a\epsilon^2}{\hat{r}^2(\hat{r}^2+\epsilon^2)^{3/2}}\;,
\nonumber\\
\kappa \hat{\widetilde{\mbox{\boldmath $T$}}}_0^{\raisebox{-.6ex}%
{\scriptsize $\;\,\alpha$}}
(\hat{x};\epsilon)&=&0\;, \; \; \; 
\kappa\hat{\widetilde{\mbox{\boldmath $T$}}}_{\alpha}^{%
\raisebox{-.6ex}{\scriptsize $\;\,0$}}
(\hat{x};\epsilon)=0\;, \nonumber \\
\kappa\hat{\widetilde{\mbox{\boldmath $T$}}}_\alpha^{%
\raisebox{-.6ex}{\scriptsize $\;\,\beta$}}
(\hat{x};\epsilon)&=&
\delta_\alpha^{\ \beta}\frac{3a\epsilon^2}{2(\hat{r}^2
+\epsilon^2)^{5/2}}-\frac{\hat{x}^\alpha \hat{x}^\beta}{\hat{r}^2}
\frac{a\epsilon^2}{(\hat{r}^2+\epsilon^2)^{5/2}}
\left(\frac{5}{2}+\frac{\epsilon^2}{\hat{r}^2}\right)\;,
\end{eqnarray}
where 
$\hat{\widetilde{\mbox{\boldmath $T$}}}_\mu^{\raisebox{-.6ex}%
{\scriptsize $\;\,\nu$}}(\hat{x};\epsilon)$
stands for the regularized 
$\hat{\widetilde{\mbox{\boldmath $T$}}}_\mu^{\raisebox{-.6ex}%
{\scriptsize $\;\,\nu$}}$.
Equations (3$\cdot $4) and (3$\cdot $5) have been obtained 
by the use of Eqs. (2$\cdot $4), (2$\cdot $12) and (2$\cdot $9) 
with Eq. (2$\cdot $10). 

From Eq. (3$\cdot$6), the desired expression 
$$
\hat{\widetilde{\mbox{\boldmath $T$}}}_\mu^{\raisebox{-.6ex}%
{\scriptsize $\;\,\nu$}}(\hat{x})
\stackrel{\mbox{\scriptsize def}}{=}
\lim_{\epsilon\to0}
\hat{\widetilde{\mbox{\boldmath $T$}}}_\mu^{\raisebox{-.6ex}%
{\scriptsize $\;\,\nu$}}
(\hat{x};\epsilon)
=-Mc^2\delta_\mu^{\ 0}\delta_0^{\ \nu}
\delta^{(3)}(\hat{\mbox{\boldmath $x$}})
\eqno{(\mbox{\ref{eq:energy-momentum-density}})}
$$ 
follows, where use has been made of the relations\footnote{
Proof of the relation (3$\cdot$8) and a comment related to it 
are given in Appendix A.} 
\begin{eqnarray}
\lim_{\epsilon\to 0}
\left[\frac{\epsilon^2}{\hat{r}^2(\hat{r}^2+\epsilon^2)^{3/2}}\right]
&=& 4\pi \delta^{(3)}(\hat{\mbox{\boldmath $x$}})\;,\nonumber \\
\lim_{\epsilon\to 0}
\left[\frac{\epsilon^2}{(\hat{r}^2+\epsilon^2)^{5/2}}\right]
&=& \frac{4}{3}\pi \delta^{(3)}(\hat{\mbox{\boldmath $x$}})\;,
\end{eqnarray}
\begin{eqnarray}
\lim_{\epsilon\to0}\left\{%
\frac{\hat{x}^\alpha \hat{x}^\beta}{\hat{r}^2}\left[%
\frac{3}{2}\frac{\epsilon^2}{(\hat{r}^2+\epsilon^2)^{5/2}}
+\frac{\epsilon^2}{\hat{r}^2(\hat{r}^2+\epsilon^2)^{3/2}}
\right]\right\}
=2\pi\delta^{\alpha \beta}\delta^{(3)}(\hat{\mbox{\boldmath $x$}})\;.
\end{eqnarray}
The regularized scalar quantities
$\hat{R}(\{\};\epsilon)$, 
$\hat{R}^{\mu\nu}(\{\};\epsilon)\hat{R}_{\mu\nu}(\{\};\epsilon)$ and 
$\hat{R}^{\rho\sigma\mu\nu}(\{\};\epsilon)
\hat{R}_{\rho\sigma\mu\nu}(\{\};\epsilon)$ are calculated as
\begin{eqnarray}
\hat{R}(\{\};\epsilon)&=&
-\frac{3a\epsilon^2}{(\hat{r}^2+\epsilon^2)^{5/2}}
+\frac{2a\epsilon^2}{\hat{r}^2(\hat{r}^2+\epsilon^2)^{3/2}}\;, 
\nonumber \\
\hat{R}^{\mu\nu}(\{\};\epsilon)\hat{R}_{\mu\nu}(\{\};\epsilon)
&=&
\frac{1}{2}\left[\frac{3a\epsilon^2}
{(\hat{r}^2+\epsilon^2)^{5/2}}\right]^2
+2\left[\frac{a\epsilon^2}
{\hat{r}^2(\hat{r}^2+\epsilon^2)^{3/2}}\right]^2 \;, 
\nonumber\\
\hat{R}^{\rho\sigma\mu\nu}(\{\};\epsilon)
\hat{R}_{\rho\sigma\mu\nu}(\{\};\epsilon)
&=&
\frac{4a^2}{\hat{r}^2+\epsilon^2}\left[%
\frac{2}{(\hat{r}^2+\epsilon^2)^2}+\frac{1}{\hat{r}^4}\right]
-\frac{12a^2\epsilon^2}{(\hat{r}^2+\epsilon^2)^4}\nonumber \\
& &+\frac{9a^2\epsilon^4}{(\hat{r}^2+\epsilon^2)^5}\;, 
\end{eqnarray}  
respectively. Thus, the scalar curvature has the 
well-defined limit
\begin{equation}
\hat{R}(\{\})\stackrel{\mbox{\scriptsize def}}{=}
\lim_{\epsilon\to0}\hat{R}(\{\};\epsilon)=
4\pi a\delta^{(3)}(\hat{\mbox{\boldmath $x$}})\;.
\end{equation}
However, the quadratic scalars do not have well-defined limits,
which can be symbolically written as
\begin{eqnarray}
\hat{R}^{\mu\nu}(\{\})\hat{R}_{\mu\nu}(\{\})
&\stackrel{\mbox{\scriptsize def}}{=}&
\lim_{\epsilon\to0}\hat{R}^{\mu\nu}(\{\};\epsilon)
\hat{R}_{\mu\nu}(\{\};\epsilon)
\sim 40\pi^2a^2\left[%
\delta^{(3)}(\hat{\mbox{\boldmath $x$}})\right]^2\;,\nonumber \\
\hat{R}^{\rho\sigma\mu\nu}(\{\})\hat{R}_{\rho\sigma\mu\nu}(\{\})
&\stackrel{\mbox{\scriptsize def}}{=}&
\lim_{\epsilon\to0}\hat{R}^{\rho\sigma\mu\nu}(\{\};\epsilon)
\hat{R}_{\rho\sigma\mu\nu}(\{\};\epsilon) \nonumber\\
&\sim& \frac{12a^2}{\hat{r}^6}
+\frac{16\pi a^2}{3}\frac{1}{\hat{r}^3}
\delta^{(3)}(\hat{\mbox{\boldmath $x$}})
+16\pi^2a^2\left[\delta^{(3)}(\hat{\mbox{\boldmath $x$}})\right]^2,
\end{eqnarray}
with $1/{\hat{r}}^{6}\stackrel{\mbox{\scriptsize def}}{=}
\lim_{\epsilon \rightarrow 0}1/(\hat{r}^{2}+\epsilon^{2})^{3}$.
Hence, we have
\begin{equation}
\hat{R}^{\mu\nu}(\{\})\hat{R}_{\mu\nu}(\{\})\neq0\;,\quad
\hat{R}^{\rho\sigma\mu\nu}(\{\})\hat{R}_{\rho\sigma\mu\nu}(\{\})
\neq \frac{12a^2}{\hat{r}^6}\;.
\end{equation}
In Eq. (3$\cdot $11), the terms 
$(1/\hat{r}^3)\delta^{(3)}(\hat{\mbox{\boldmath $x$}})$
and 
$\left[\delta^{(3)}(\hat{\mbox{\boldmath $x$}})\right]^2$
are both ill-defined, and the symbol $\sim $ is to denote the 
\lq\lq equality" of the left- and right-hand sides in a rough sense.
Mathematical formalism capable of describing these singular quantities
is needed.

It is worth mentioning here that the spherically symmetric non-scalar
quantities\linebreak
$\hat{r}^{3}\hat{R}^{\mu\nu}(\{\};\epsilon)\hat{R}_{\mu\nu}(\{\};\epsilon)$
and $\hat{r}^{3}\hat{R}^{\rho\sigma\mu\nu}(\{\};\epsilon)
\hat{R}_{\rho\sigma\mu\nu}(\{\};\epsilon)$ have the well-defined
limits
\begin{eqnarray}
  \lim_{\epsilon\to0}\hat{r}^{3}
  \hat{R}^{\mu\nu}(\{\};\epsilon)\hat{R}_{\mu\nu}(\{\};\epsilon)
  &=&\frac{11}{4}\pi a^{2}\delta^{(3)}(\hat{\mbox{\boldmath $x$}})\;, 
  \nonumber \\
  \lim_{\epsilon\to0}\hat{r}^{3}
  \hat{R}^{\rho\sigma\mu\nu}(\{\};\epsilon)
  \hat{R}_{\rho\sigma\mu\nu}(\{\};\epsilon)
  &=&12a^{2}\left[\frac{1}{\hat{r}^{3}}\right]
  +\frac{11}{2}\pi a^{2}\delta^{(3)}(\hat{\mbox{\boldmath $x$}})
\end{eqnarray}
with 
\begin{equation}
  \left[\frac{1}{{\hat{r}}^{3}}\right]
  \stackrel{\mbox{\scriptsize def}}{=}
  \lim_{\epsilon \rightarrow 0}
  \frac{\hat{r}^{3}}{(\hat{r}^{2}+\epsilon^{2})^{3}}\;.
\end{equation}

\section{Generalization}
\label{sec:4}

\hspace*{3ex} We now consider the problem in the coordinates
$(x^{0},r,\theta, \phi )$ related to the Schwarzschild coordinates
$(\hat{x}^{0}, \hat{r}, \theta, \phi)$ through the relation (1$\cdot
$4). In the coordinates $(x^{0},r,\theta,\phi)$, the metric has the
form
\begin{eqnarray}
  ds^2&=&-A(dx^0)^2-2Ddx^0dr+(B+C)(dr)^2\nonumber \\
  & &+Br^2\left[%
    (d\theta)^2+\sin^2\theta(d\phi)^2\right]\;,
\end{eqnarray}
where we have defined 
\begin{eqnarray}
A &\stackrel{\mbox{\scriptsize def}}{=}&
\Omega^2\left(1-\frac{a}{K}\right)\;,\quad 
B \stackrel{\mbox{\scriptsize def}}{=}\frac{K^2}{\rho^2}\;,
\nonumber\\
C &\stackrel{\mbox{\scriptsize def}}{=}&
\left(1-\frac{a}{K}\right)^{-1}(K')^2-\frac{K^2}{\rho^2}
-\left(1-\frac{a}{K}\right)(f')^2\;,\nonumber \\ 
D &\stackrel{\mbox{\scriptsize def}}{=}&
\Omega\left(1-\frac{a}{K}\right)f'
\end{eqnarray}
with\footnote{The manipulation to denote $r$ with $\rho$ is for the 
regularization procedure employed below.}
 $\rho \stackrel{\mbox{\scriptsize def}}{=}r$, 
$K' \stackrel{\mbox{\scriptsize def}}{=} dK/dr$ and $f' 
\stackrel{\mbox{\scriptsize def}}{=} df/dr$. 
The coordinates $t=x^{0}/c$ and $r$ are time and space coordinates,
respectively, only if 
\begin{equation}
1-\frac{a}{K}>0\;, \; \; \; 
\left(1-\frac{a}{K}\right)^{-1}(K')^{2}
-\left(1-\frac{a}{K}\right)(f')^{2}>0\;. 
\end{equation}
In the Cartesian coordinate system $\{x^{\mu }; \mu =0,1,2,3\}$ with
$x^{1}=r\cos\phi\sin\theta$, $x^{2}$\linebreak$=r\sin\phi\sin\theta$,
$x^{3}=r\cos\theta$, the metric takes the form
\begin{equation}
ds^{2}=g_{\mu \nu }dx^{\mu }dx^{\nu }
\end{equation}
with $g_{\mu \nu }$ given by
\begin{equation}
g_{00}=-A\;, \; \; g_{0 \alpha }=-D\frac{x^{\alpha }}{r}\;, \; \;
g_{\alpha \beta}=B\delta^{\alpha \beta}
+C\frac{x^{\alpha }x^{\beta }}{r^{2}}\;.
\end{equation}

We can obtain expressions for the quantities 
$\widetilde{\mbox{\boldmath $t$}}_\mu^{\ \nu}$, 
$\widetilde{\mbox{\boldmath $T$}}_\mu^{\ \nu}$,
$R(\{\})$, $R^{\mu\nu}(\{\})R_{\mu\nu}(\{\})$ and \\
$R^{\rho\sigma\mu\nu}(\{\})R_{\rho\sigma\mu\nu}(\{\})$ in
terms of $A,B,C,D$, which are enumerated in 
Appendix B. 

By using Eqs. (4$\cdot $2), (B$\cdot $1) and (B$\cdot $2), 
we can show that 
\begin{equation}
\widetilde{\mbox{\boldmath $t$}}_\mu^{\ \nu}=0\;, 
\end{equation}
if {\bf Case 1}:
\begin{equation}
K(r)=\Lambda r\;, 
\end{equation}
or {\bf Case 2}:
\begin{equation}
K(r)=\Lambda r+a\;, \; \; f(r)=\Gamma r+\Xi  
\end{equation}
with $\Lambda ,\Gamma $ and $\Xi $ being real constants. 
(For {\bf Case 1}, $f(r)$ is arbitrary.)

In what follows, we restrict our consideration to the above 
two cases, {\bf Case 1} and {\bf Case 2} with 
positive\footnote{Note the following: (a) It is expected that a 
delta function expression for 
$\widetilde{\mbox{\boldmath $T$}}_\mu^{\ \nu}(x)$ is obtained when 
$\widetilde{\mbox{\boldmath $t$}}_\mu^{\ \nu}=0\;$. 
(b) The variable $r$ cannot be \lq\lq\emph{radial\/}"
if $\Lambda \leq 0$.\label{footnote:auxiliary}} $\Lambda$.
In order to treat the singularity at $r=0$ in a distribution 
theoretical way, we employ the following 
regularization scheme:
\begin{description}
\item[R.1]\ First, replace $K'$ with $\Lambda $ in the expression 
of the function $C$.
\item[R.2]\ Then, replace $K'$ and $K''$ appearing in the process of
calculation with $\Lambda \rho'$ and $\Lambda \rho''$, respectively.
\item[R.3]\ Finally, replace the function $\rho$ 
with $\sqrt{r^{2}+\epsilon^{2}}$. 
\end{description}
Then, we obtain the following results by using Eqs. (4$\cdot $2), 
(B$\cdot $1)$\sim $(B$\cdot $4):\\
{\bf Case 1}
\begin{description}
\item[(1)] The regularized energy-momentum density 
$\widetilde{\mbox{\boldmath $T$}}_\mu^{\ \nu}(x;\epsilon)$ is 
given by 
\begin{eqnarray}
\kappa\widetilde{\mbox{\boldmath $T$}}_0^{\ 0}(x;\epsilon)
&=& -|\Omega|\frac{a\epsilon^2}{r^2(r^2+\epsilon^2)^{3/2}}\;,
\nonumber \\
\kappa\widetilde{\mbox{\boldmath $T$}}_0^{\ \alpha}(x;\epsilon)&=&
0\;,\; \; \;
\kappa\widetilde{\mbox{\boldmath $T$}}_\alpha^{\ 0}(x;\epsilon)=0\;,
\nonumber \\
\kappa\widetilde{\mbox{\boldmath $T$}}_\alpha^{\ \beta}(x;\epsilon)
&=&|\Omega|\left[%
\delta_\alpha^{\ \beta}\frac{3a\epsilon^2}{2(r^2+\epsilon^2)^{5/2}}
-\frac{x^\alpha x^\beta}{r^2}
\frac{a\epsilon^2}{(r^2+\epsilon^2)^{5/2}}
\left(\frac{5}{2}+\frac{\epsilon^2}{r^2}\right)\right]\;.\nonumber \\
\end{eqnarray}
This leads to 
\begin{equation}
\widetilde{\mbox{\boldmath $T$}}_\mu^{\ \nu}(x)
\stackrel{\mbox{\scriptsize def}}{=}
\lim_{\epsilon\to0}
\widetilde{\mbox{\boldmath $T$}}_\mu^{\ \nu}(x;\epsilon)
=-Mc^2|\Omega|
\delta_\mu^{\ 0}\delta_0^{\ \nu}
\delta^{(3)}(\mbox{\boldmath $x$})\;,
\end{equation}
which reduces to
\begin{equation}
\widetilde{\mbox{\boldmath $T$}}_\mu^{\ \nu}(x)
=-Mc^2\delta_\mu^{\ 0}\delta_0^{\ \nu}
\delta^{(3)}(\mbox{\boldmath $x$})
\end{equation}
when $|\Omega |=1$. 
There is the relation
\begin{equation} 
\widetilde{\mbox{\boldmath $T$}}_{\mu }^{\ \nu }(x;\epsilon)
=\left[\widetilde{\mbox{\boldmath $T$}}_{\mu }^{\ \nu }
(x;\epsilon/\Lambda)\right]_{c}\;,
\end{equation}
where we have defined
\begin{equation}
\left[\widetilde{\mbox{\boldmath $T$}}_{\mu }^{\ \nu }
(x;\epsilon)\right]_{c}
\stackrel{\mbox{\scriptsize def}}{=}
\frac{\partial (\hat{x})}{\partial (x)}
\frac{\partial \hat{x}^{\lambda }}{\partial x^{\mu }}
\frac{\partial x^{\nu }}{\partial \hat{x}^{\rho }}
\hat{\widetilde{\mbox{\boldmath $T$}}}_\lambda^{\raisebox{-.6ex}%
{\scriptsize $\;\,\rho$}}
(\hat{x};\epsilon)\;.
\end{equation}
Thus, we see that 
\begin{equation} 
\widetilde{\mbox{\boldmath $T$}}_\mu^{\ \nu}(x)=
\lim_{\epsilon\to0}
\left[\widetilde{\mbox{\boldmath $T$}}_{\mu }^{\ \nu }
(x;\epsilon)\right]_{c}\;.
\end{equation}
For 
\begin{equation}
\left[\widetilde{\mbox{\boldmath $T$}}_\mu^{\ \nu}(x)\right]_{c}
\stackrel{\mbox{\scriptsize def}}{=}
\frac{\partial (\hat{x})}{\partial (x)}
\frac{\partial \hat{x}^{\lambda }}{\partial x^{\mu }}
\frac{\partial x^{\nu }}{\partial \hat{x}^{\rho }}
\hat{\widetilde{\mbox{\boldmath $T$}}}_\lambda^{\raisebox{-.6ex}%
{\scriptsize $\;\,\rho$}}(\hat{x})\;,
\end{equation}
however, we have 
\begin{eqnarray}
\left[\widetilde{\mbox{\boldmath $T$}}_0^{\ 0}(x)\right]_{c}
&=&-Mc^{2}|\Omega|
\delta^{(3)}(\mbox{\boldmath $x$})\;,\; \; 
\left[\widetilde{\mbox{\boldmath $T$}}_0^{\ \alpha}(x)
\right]_{c}=0\;,\nonumber \\
\noalign{\vskip 1ex}
\left[\widetilde{\mbox{\boldmath $T$}}_\alpha^{\ 0}(x)\right]_{c}
&=&-\frac{Mc^{2}|\Omega |}{\Omega}f'(r)\frac{x^{\alpha}}{r}
\delta^{(3)}(\mbox{\boldmath $x$})\;, \nonumber \\
\noalign{\vskip 1ex}
\left[\widetilde{\mbox{\boldmath $T$}}_\alpha^{\ \beta}(x)
\right]_{c}&=&0\;,
\end{eqnarray}
in which the ill-defined quantity\footnote{%
\label{footnote:ill-defined}Note that 
$x^{\alpha }[(1/r)\delta^{(3)}(\mbox{\boldmath $x$})]\neq 
0=(1/r)[x^{\alpha }\delta^{(3)}(\mbox{\boldmath $x$})]$ and 
that $\left[\sum_{\alpha =1}^{3}
\frac{x^{\alpha }}{r}\frac{x^{\alpha }}{r}\right]
\delta^{(3)}(\mbox{\boldmath $x$})=
\delta^{(3)}(\mbox{\boldmath $x$})$. See also the relation 
(A$\cdot $16) in Appendix A.} 
$\frac{x^{\alpha }}{r}\delta^{(3)}(\mbox{\boldmath $x$})$ appears, 
and hence 
\begin{equation}
\widetilde{\mbox{\boldmath $T$}}_{\mu }^{\ \nu}(x)
\neq 
\left[\widetilde{\mbox{\boldmath $T$}}_\mu^{\ \nu}(x)
\right]_{c}\;.
\end{equation}
From Eqs. (4$\cdot $14) and (4$\cdot $17), we see the following: 
The correct expression for the energy-momentum density 
$\widetilde{\mbox{\boldmath $T$}}_{\mu }^{\ \nu}(x)$ is obtainable by
transforming {\em first} the regularized density 
$\hat{\widetilde{\mbox{\boldmath $T$}}}_\mu^{\raisebox{-.6ex}%
{\scriptsize $\;\,\nu$}}(\hat{x};\epsilon)$ and
{\em then} taking the limit $\epsilon \rightarrow 0$. 
However, it cannot be obtained, 
if the order of making coordinate transformation and taking 
the limit is exchanged.

In Ref.~\cite{Balasin2}, the expression 
$\widetilde{\mbox{\boldmath $T$}}_\mu^{\ \nu}(x)
=-Mc^2\delta_\mu^{\ 0}\delta_0^{\ \nu}
\delta^{(3)}(\mbox{\boldmath $x$})$
is given for Kerr-Schild coordinates. One might think that 
Eq. (1$\cdot $3) can be easily obtained by transforming this to 
Schwarzschild coordinates. However, this is not the case;
the situation here is similar to the one discussed above.
\item[(2)]For the regularized scalar quantities, we have
\begin{eqnarray}
R(\{\};\epsilon)&=&\Lambda^{-3}\left[
-\frac{3a\epsilon^2}{(r^2+\epsilon^2)^{5/2}}
+\frac{2a\epsilon^2}{r^2(r^2+\epsilon^2)^{3/2}}\right]
=\hat{R}(\{\};\Lambda \epsilon)\;, 
\nonumber \\
R^{\mu\nu}(\{\};\epsilon)R_{\mu\nu}(\{\};\epsilon)
&=&\Lambda^{-6}\left\{
\frac{1}{2}\left[\frac{3a\epsilon^2}{(r^2+\epsilon^2)^{5/2}}\right]^2
+2\left[\frac{a\epsilon^2}
{r^2(r^2+\epsilon^2)^{3/2}}\right]^2\right\}\nonumber \\
&=&\hat{R}^{\mu\nu}(\{\};\Lambda \epsilon)
\hat{R}_{\mu\nu}(\{\};\Lambda \epsilon)\;,
\nonumber \\
R^{\rho\sigma\mu\nu}(\{\};\epsilon)
R_{\rho\sigma\mu\nu}(\{\};\epsilon)
&=&\Lambda^{-6}\left\{%
\frac{4a^2}{r^2+\epsilon^2}\left[%
\frac{2}{(r^2+\epsilon^2)^2}+\frac{1}{r^4}\right]
-\frac{12a^2\epsilon^2}{(r^2+\epsilon^2)^4}\right.\nonumber \\
& &\qquad\quad+\left.\frac{9a^2\epsilon^4}{(r^2+\epsilon^2)^5}\right\}
\nonumber \\
&=&\hat{R}^{\rho\sigma\mu\nu}(\{\};\Lambda \epsilon)
\hat{R}_{\rho\sigma\mu\nu}(\{\};\Lambda \epsilon)\;.
\end{eqnarray}
Equation (4$\cdot$18) implies that the regularized scalars in the
coordinate system $\{x^{\mu};\mu$ \linebreak $=0,1,2,3\}$ for {\bf
  Case 1} essentially agree with the corresponding scalars in
Schwarzschild coordinates.\footnote{Note that the product $\Lambda
  \epsilon $ in the hatted scalars in Eq. (4$\cdot $18) plays
  essentially the same role as $\epsilon $, because the limit
  $\epsilon \rightarrow 0$ is taken in the final stages.}  This is not
self-evident, because we have made the regularization in each
coordinate system independently.\footnote{See also the paragraph at
  the end of this section.}
\end{description}
{\bf Case 2}
\begin{description}
\item[(1)] The regularized energy-momentum density 
$\widetilde{\mbox{\boldmath $T$}}_\mu^{\ \nu}(x;\epsilon)$ 
is given by
\begin{eqnarray}
\kappa\widetilde{\mbox{\boldmath $T$}}_0^{\ 0}(x;\epsilon)
&=&|\Omega|\left[%
-\frac{6a\epsilon^2}{(r^2+\epsilon^2)^{5/2}}
+\frac{a\epsilon^2}{r^2(r^2+\epsilon^2)^{3/2}}\right]\;, \nonumber \\
\kappa\widetilde{\mbox{\boldmath $T$}}_0^{\ \alpha}(x;\epsilon)&=&0\;,
\; \; \;
\kappa\widetilde{\mbox{\boldmath $T$}}_\alpha^{\ 0}(x;\epsilon)
=-\frac{|\Omega|}{\Omega}\Gamma\frac{x^\alpha}{r}
\left[\frac{6a\epsilon^2}{(r^2+\epsilon^2)^{5/2}}\right]\;, \nonumber\\
\kappa\widetilde{\mbox{\boldmath $T$}}_\alpha^{\ \beta}(x;\epsilon)
&=&|\Omega|\left[%
-\delta_\alpha^{\ \beta}\frac{3a\epsilon^2}{2(r^2+\epsilon^2)^{5/2}}
+\frac{x^\alpha x^\beta}{r^2}
\frac{a\epsilon^2}{(r^2+\epsilon^2)^{5/2}}
\left(\frac{5}{2}+\frac{\epsilon^2}{r^2}\right)\right]\;,
\nonumber \\
\end{eqnarray}
which yields a limit having the same form as Eq. (4$\cdot $10), 
upon using Eqs. (3$\cdot $7), (3$\cdot $8) 
and the relation\footnote{Proof of 
the relation (4$\cdot $20) and a comment related to it are given 
in Appendix A.}
\begin{equation}
\lim_{\epsilon\to0}\left[%
\frac{x^\alpha}{r}\frac{\epsilon^2}{(r^2+\epsilon^2)^{5/2}}
\right]=0\;.
\end{equation}

The quantity 
$\left[\widetilde{\mbox{\boldmath $T$}}_\mu^{\ \nu}
(x;\epsilon)\right]_{c}$ 
defined in the same way as Eq. (4$\cdot $13) is given by 
\begin{eqnarray}
\kappa\left[\widetilde{\mbox{\boldmath $T$}}_0^{\ 0}
(x;\epsilon)\right]_{c}
&=&-|\Lambda \Omega |\frac{a\epsilon^{2}}{r^{2}
[(\Lambda r+a)^{2}+\epsilon^{2}]^{3/2}}\;, \nonumber \\
\kappa\left[\widetilde{\mbox{\boldmath $T$}}_0^{\ \alpha}
(x;\epsilon)\right]_{c} &=& 0\;, \; \; 
\kappa\left[\widetilde{\mbox{\boldmath $T$}}_\alpha^{\ 0}
(x;\epsilon)\right]_{c}=0\;, \nonumber \\
\kappa\left[\widetilde{\mbox{\boldmath $T$}}_\alpha^{\ \beta}
(x;\epsilon)\right]_{c}
&=&|\Lambda \Omega|\left(\frac{\Lambda r +a}{r}\right)^{2}
\left\{\frac{3a}{2}
\frac{\epsilon^{2}}{[(\Lambda r+a)^{2}+\epsilon^{2}]^{5/2}}
\delta_{\alpha}^{\beta}\right. \nonumber \\
& &\left. -\frac{a\epsilon^{2}}
{[(\Lambda r+a)^{2}+\epsilon^{2}]^{3/2}}
\frac{x^{\alpha }x^{\beta }}{r^{2}}
\left[\frac{1}{(\Lambda r+a)^{2}}
+\frac{3}{2}\frac{1}{(\Lambda r+a)^{2}+\epsilon^{2}}\right]\right\}\;,
\nonumber \\
\end{eqnarray}
from which the limit
\begin{equation}
\lim_{\epsilon\to0}
\left[\widetilde{\mbox{\boldmath $T$}}_\mu^{\ \nu}
(x;\epsilon)\right]_{c}=0
\neq \widetilde{\mbox{\boldmath $T$}}_\mu^{\ \nu}(x)
\end{equation}
is obtained. Also, 
$\left[\widetilde{\mbox{\boldmath $T$}}_\mu^{\ \nu}
(x)\right]_{c}$ defined
in the same way as for {\bf Case 1} has the expression
\begin{eqnarray}
\left[\widetilde{\mbox{\boldmath $T$}}_0^{\ 0}
(x)\right]_{c}
&=&-Mc^{2}|\Lambda \Omega |\left(\frac{\Lambda r+a}{r}\right)^{2}
\delta^{(3)}(\Lambda \mbox{\boldmath $x$}+\mbox{\boldmath $a$})\;, 
\nonumber \\
\noalign{\vskip 1ex} 
\left[\widetilde{\mbox{\boldmath $T$}}_0^{\ \alpha}(x)
\right]_{c}&=&0\;, \; \;
\left[\widetilde{\mbox{\boldmath $T$}}_\alpha^{\ \beta}(x)
\right]_{c}=0\;, 
\nonumber \\
\noalign{\vskip 1ex}
\left[\widetilde{\mbox{\boldmath $T$}}_\alpha^{\ 0}(x)
\right]_{c}
&=&-Mc^{2}\frac{{|\Lambda \Omega|}}{\Omega }
\Gamma \left(\frac{\Lambda r+a}{r}\right)^{2}\frac{x^{\alpha}}{r}
\delta^{(3)}(\Lambda \mbox{\boldmath $x$}+\mbox{\boldmath $a$})
\nonumber \\
\end{eqnarray}
with $\mbox{\boldmath $a$}\stackrel{\mbox{\scriptsize def}}{=}
a \mbox{\boldmath $x$}/r$, and we have a relation having the same 
form as Eq. (4$\cdot $17). 
This and Eq. (4$\cdot $22) imply that the correct expression for 
$\widetilde{\mbox{\boldmath $T$}}_{\mu }^{\ \nu}(x)$ is {\em not} 
obtainable for {\bf Case 2} by applying the coordinate 
transformation and the limiting procedure to the regularized density
$\hat{\widetilde{\mbox{\boldmath $T$}}}_\mu^{\raisebox{-.6ex}%
{\scriptsize $\;\,\nu$}}(\hat{x};\epsilon)$. 
\item[(2)] The regularized scalar quantities are given by 
\begin{eqnarray}
R(\{\};\epsilon)&=&
\frac{r^2+\epsilon^2}{\Lambda(\Lambda\sqrt{r^2+\epsilon^2}+a)^2}
\left[\frac{9a\epsilon^2}{(r^2+\epsilon^2)^{5/2}}
-\frac{2a\epsilon^2}{r^2(r^2+\epsilon^2)^{3/2}}\right]
\;, \nonumber \\
R^{\mu\nu}(\{\};\epsilon)R_{\mu\nu}(\{\};\epsilon) 
&=&\frac{(r^2+\epsilon^2)^2}{\Lambda^2
(\Lambda\sqrt{r^2+\epsilon^2}+a)^4}
\left\{\frac{5}{2}
\left[\frac{3a\epsilon^2}{(r^2+\epsilon^2)^{5/2}}\right]^2 
\right. \nonumber \\
&&\left. +2\left[-\frac{3a\epsilon^2}{(r^2+\epsilon^2)^{5/2}}
+\frac{a\epsilon^2}{r^2(r^2+\epsilon^2)^{3/2}}\right]^{2}\right\} 
\;,\nonumber \\
R^{\rho\sigma\mu\nu}(\{\};\epsilon)R_{\rho\sigma\mu\nu}
(\{\};\epsilon) 
&=&\frac{12a^2}{\Lambda^2(\Lambda \sqrt{r^2+\epsilon^2}+a)^6}
\left[\Lambda+\frac{a\epsilon^2}{(r^2+\epsilon^2)^{3/2}}\right]^2\nonumber\\
&&-\frac{4a^2}{\Lambda^2(\Lambda\sqrt{r^2+\epsilon^2}+a)^5}
\left[\Lambda+\frac{a\epsilon^2}{(r^2+\epsilon^2)^{3/2}}\right]\nonumber\\
&&\quad\times\left[\frac{2\epsilon^2}{r^2\sqrt{r^2+\epsilon^2}}
+\frac{9\epsilon^2}{(r^2+\epsilon^2)^{3/2}}\right]\nonumber \\
&&+\frac{a^2}{\Lambda^2(\Lambda\sqrt{r^2+\epsilon^2}+a)^4}
\left[\frac{4\epsilon^4}{r^4(r^2+\epsilon^2)}
+\frac{81\epsilon^4}{(r^2+\epsilon^2)^3}\right]\;,\nonumber \\
\end{eqnarray}
from which the following is known:
\begin{eqnarray}
R(\{\})&\stackrel{\mbox{\scriptsize def}}{=}&
\lim_{\epsilon\to0}R(\{\};\epsilon)=0\;,\nonumber \\
R^{\mu\nu}(\{\};\epsilon)R_{\mu\nu}(\{\};\epsilon)
&\sim &\frac{2a^2}{\Lambda^2(\Lambda r+a)^4}
\frac{\epsilon^4}{r^4(r^2+\epsilon^2)}\neq0\;,
\nonumber \\
R^{\rho\sigma\mu\nu}(\{\};\epsilon)
R_{\rho\sigma\mu\nu}(\{\};\epsilon)
&\sim &\frac{12a^2}{(\Lambda r+a)^6}+
\frac{4a^2}{\Lambda^2(\Lambda r+a)^4}
\frac{\epsilon^4}{r^4(r^2+\epsilon^2)}\nonumber \\
&\neq &\frac{12a^{2}}{(\Lambda r+a)^{6}}\;.
\end{eqnarray}
Also, we can show that
\begin{eqnarray}
\lim_{\epsilon\to0}r^{\omega}
R^{\mu\nu}(\{\};\epsilon)R_{\mu\nu}(\{\};\epsilon)
&=&0\;, \nonumber \\
\lim_{\epsilon\to0}r^{\omega}
R^{\rho\sigma\mu\nu}(\{\};\epsilon)
R_{\rho\sigma\mu\nu}(\{\};\epsilon)&=&0\;,
\; \; {\rm for}\; \; \omega >1\;.
\end{eqnarray}
As is known from Eqs. (3$\cdot $9) and (4$\cdot $24), 
the regularized scalars in the coordinate system 
$\{x^{\mu};\mu =0,1,2,3\}$ for 
{\bf Case 2} are not simply related to the corresponding scalars in 
Schwarzschild coordinates. However, this is {\em not} a contradiction, 
because the regularization has been made in each coordinate system 
independently.
\end{description}

\section{Summary and comments}
\label{sec:5}

\hspace*{3ex} In the above, we have examined the Schwarzschild
space-time from a distribution theoretical point of view. Expressions
of the energy-momentum densities $\widetilde{\mbox{\boldmath
    $T$}}_\mu^{\ \nu}$ and $\widetilde{\mbox{\boldmath $t$}}_\mu^{\ 
  \nu}$ and of the scalars $R(\{\}), R^{\mu\nu}(\{\})R_{\mu\nu}(\{\})$
and $R^{\rho\sigma\mu\nu}(\{\}) R_{\rho\sigma\mu\nu}(\{\})$ have been
obtained for a class of coordinate systems including those of the
Schwarzschild and of Kerr-Schild types as special cases.

The results can be summarized as follows: 
\begin{description}
\item[(1)] For Schwarzschild coordinates, employing the regularization
  scheme prescribed by the replacement $a/\hat{r}$ with
  $a/\sqrt{\hat{r}^{2}+\epsilon^{2}}$ in the expression (3$\cdot $3)
  of the components of the metric, we have shown the following:
\begin{description}
\item[(1.A)]\ The energy-momentum density 
$\hat{\widetilde{\mbox{\boldmath $T$}}}_\mu^{\raisebox{-.6ex}%
  {\scriptsize $\;\,\nu$}}(\hat{x})$ of the gravitational source has
the expression (1$\cdot $3).
\item[(1.B)]\ The scalar $\hat{R}(\{\})$ has the definite delta
  function expression (3$\cdot $10), while
  $\hat{R}^{\mu\nu}(\{\})\hat{R}_{\mu\nu}(\{\})$ and
  $\hat{R}^{\rho\sigma\mu\nu}(\{\})\hat{R}_{\rho\sigma\mu\nu}(\{\})$
  have the expression (3$\cdot $11), which include terms like
  $\left[\delta^{(3)}(\mbox{\boldmath $x$)}\right]^2$ and
  $\delta^{(3)}(\mbox{\boldmath $x$})/r^3$. Also,
  $\hat{r}^{3}\hat{R}^{\mu\nu}(\{\};\epsilon)
  \hat{R}_{\mu\nu}(\{\};\epsilon)$ and
  $\hat{r}^{3}\hat{R}^{\rho\sigma\mu\nu}(\{\};\epsilon)
  \hat{R}_{\rho\sigma\mu\nu}(\{\};\epsilon)$ have the definite limits
  (3$\cdot $13).  The second relation in Eq. (3$\cdot$12) shows that
  the well-known expression $\hat{R}^{\rho\sigma\mu\nu}(\{\})
  \hat{R}_{\rho\sigma\mu\nu}(\{\})=12a^{2}/\hat{r}^{6}$ is not
  correct, as long as we define
  $1/\hat{r}^6\stackrel{\mbox{\scriptsize def}}{=}
  \lim_{\epsilon\to0}1/(\hat{r}^2+\epsilon^2)^3\;$.
\end{description}
\item[(2)]For the coordinates $(x^{0}, r, \theta, \phi )$ related to
  the Schwarzschild coordinates $(\hat{x}^{0},\hat{r}, \theta, \phi )$
  through (1.4), we have shown the following:
\begin{description}
\item[(2.A)]\ If {\bf Case 1}: $K(r)=\Lambda r$, or {\bf Case 2}:
  $K(r)=\Lambda r+a\;, f(r)=\Gamma r+\Xi $ with $\Lambda , \Gamma $
  and $\Xi $ being real constants, the gravitational energy-momentum
  density vanishes, $\widetilde{\mbox{\boldmath $t$}}_\mu^{\ \nu }=0$.
\item[(2.B)]\ Restricting ourselves to cases with positive $\Lambda $
  and employing the regularization scheme prescribed by {\bf R.1},
  {\bf R.2} and {\bf R.3}, we have shown the
  following:\\
  For both {\bf Case 1} and {\bf Case 2}, the energy-momentum density
  $\widetilde{\mbox{\boldmath $T$}}_\mu^{\ \nu}(x)$ has the delta
  function expression (4$\cdot $10), which reduces to
  $\widetilde{\mbox{\boldmath $T$}}_\mu^{\ \nu}(x) =-Mc^2\delta_\mu^{\ 
    0}\delta_0^{\ \nu}
  \delta^{(3)}(\mbox{\boldmath $x$})$, when $|\Omega |=1$.\\
  {\bf Case 1}
\begin{description}
\item[(2.1.a)]\ The correct expression for $\widetilde{\mbox{\boldmath
      $T$}}_{\mu }^{\ \nu}(x)$ can be obtained by transforming {\em
    first} the regularized density
$\hat{\widetilde{\mbox{\boldmath $T$}}}_\mu^{\raisebox{-.6ex}%
  {\scriptsize $\;\,\nu$}}(\hat{x};\epsilon)$ and {\em then} taking
the limit $\epsilon \rightarrow 0$.  However, it cannot be obtained if
the order of the coordinate transformation and the limiting procedure
is exchanged.
\item[(2.1.b)]\ The regularized scalars $R(\{\};\epsilon),$
  $R^{\mu\nu}(\{\};\epsilon)R_{\mu\nu}(\{\};\epsilon)$ and
  $R^{\rho\sigma\mu\nu}(\{\};\epsilon)$\linebreak $\times
  R_{\rho\sigma\mu\nu}(\{\};\epsilon)$ are essentially equal to the
  corresponding scalars in Schwarzschild coordinates.
\end{description}
{\bf Case 2}
\begin{description}
\item[(2.2.a)]\ The correct expression for $\widetilde{\mbox{\boldmath
      $T$}}_\mu^{\ \nu}(x)$ cannot be obtained by considering the
  transformation of the regularized or limiting densities in
  Schwarzschild coordinates.
\item[(2.2.b)]\ The regularized scalars $R(\{\};\epsilon)$,
  $R^{\mu\nu}(\{\};\epsilon)R_{\mu\nu}(\{\};\epsilon)$ and
  $R^{\rho\sigma\mu\nu}(\{\};\epsilon)$\linebreak $\times
  R_{\rho\sigma\mu\nu}(\{\};\epsilon)$ are given by Eq. (4$\cdot $24),
  from which Eqs. (4$\cdot $25) and (4$\cdot $26) follow.
\end{description}
\end{description}
\end{description}

The following is worth mentioning:
\begin{description}
\item[{[A]}]\ When the condition $|\Omega |=1$ is satisfied, the
  gravity appears to be interpreted as produced by a point-like
  particle of mass $M$ located at $\mbox{\boldmath
    $x$}=\mbox{\boldmath $0$}$ for both {\bf Case 1} and {\bf Case 2},
  so far as energy-momentum densities are concerned.  Also for the
  orbit of a test point-like particle moving in the region $r>a$, this
  interpretation works well for
{\bf Case 1} if the additional condition\footnote{%
  Note that the condition (\ref{eq:additional-condition}) is satisfied
  for Schwarzschild and Kerr-Schild coordinates.}
\begin{equation}
  \label{eq:additional-condition}
  |f'|\left(1-\frac{a}{r}\right)<1\;,
\end{equation}
following from Eq. (4$\cdot $3), is satisfied. But, for {\bf Case 2},
a test point-like particle is expected to move as if the source were
located at $r=-a/\Lambda $.  The fact that energy-momentum density
$\widetilde{\mbox{\boldmath $T$}}_\mu^{\raisebox{-.6ex}%
  {\scriptsize $\;\,\nu$}}(x)$ takes the form (4$\cdot $11) does not
necessarily imply that the gravity is produced by a mass point located
at $\mbox{\boldmath $x$}=\mbox{\boldmath $0$}$.
\item[{[B]}]\ The situation is simple and natural in {\bf Case 1}, as
  in Schwarzschild coordinates. This is reasonable, because the
  Jacobian of the transformation (\ref{eq:coordinate-transformation})
  for this case is given by
\begin{equation}
  \frac{\partial(x')}{\partial(x)}=\Omega\Lambda^3\;,
\end{equation}
which implies that the transformation is everywhere regular.
\item[{[C]}]\ The volume element $\sqrt{-g}dx^{0}dx^{1}dx^{2}dx^{3}$ 
is given by 
\begin{equation}
\sqrt{-g}dx^{0}dx^{1}dx^{2}dx^{3}
=|\Omega \Lambda^{3}|dx^{0}dx^{1}dx^{2}dx^{3}\;, 
\end{equation}
for {\bf Case 1}, which includes Schwarzschild coordinates and
Kerr-Schild coordinates as special cases. This case describes
volume-preserving coordinates, which are preferred in the sense
mentioned in Ref.~\cite{Zalalet}.
\item[{[D]}]\ For {\bf Case 2}, the situation is rather curious, as is
  known from the statement {\bf [A]}. The following is worth noting in
  this connection: The Jacobian of the coordinate transformation
  (\ref{eq:coordinate-transformation}) for this case is
\begin{equation}
  \frac{\partial(x')}{\partial(x)}
  =\Omega \Lambda \frac{(\Lambda r+a)^2}{r^2}\;, 
\end{equation}
which means that the coordinate transformation is singular at $r=0$
and at $\Lambda r$\linebreak$+a=0$. Thus, the coordinate systems of
{\bf Case 2} and of Schwarzschild are not equivalent, and the
corresponding metrics describe different physical situations.$^{6)}$
\item[{[E]}]\ The singularity at $\mbox{\boldmath $x$}
  =\mbox{\boldmath $0$}$ in {\bf Case 2} is weaker than in the case of
  the Schwarzschild coordinate system and in {\bf Case 1}, as is seen
  from Eqs. (3$\cdot $10), (3$\cdot $11), (3$\cdot $13), (4$\cdot
  $18), (4$\cdot $25) and (4$\cdot $26). This is expected from the
  regularity and singularity of the Jacobians (5$\cdot $2) and
  (5$\cdot $4).
\item[{[F]}]\ We must be careful in considering the coordinate
  transformation of distributional quantities, as is known from Eqs.
  (3$\cdot $9), (3$\cdot$10), (3$\cdot$11), (4$\cdot$24), (4$\cdot$25)
  and the statements {\bf (2.1.a)} and {\bf (2.2.a)}.
\item[{[G]}]\ The regularization schemes employed in $\S\S$ 3 and 4
  are crucial in giving our results.  We have regularized the \lq\lq
  dynamical part" only. If we had regularized the \lq\lq kinematical
  factors" $1/\hat{r}$ and $1/\hat{r}^{2}$ in Eq. (3$\cdot $4) as
  $1/\sqrt{\hat{r}^{2}+\epsilon^{2}}$ and
  $1/(\hat{r}^{2}+\epsilon^{2})$, respectively, in addition to
  regularizing the \lq\lq dynamical function" $h=a/\hat{r}$, for
  example, we would not have obtained Eq.
  (\ref{eq:energy-momentum-density}).  Also, if we had not followed
  the scheme prescribed by {\bf R.1}, {\bf R.2} and {\bf R.3} in
  $\S$4, we might have been led to unwelcome results. However, as for
  this scheme, we cannot simply say that we have regularized \lq \lq
  dynamical parts" only.  At present, we cannot give a lucid
  interpretation to the scheme employed in $\S$4.
\item[{[H]}]\ Sometimes, a singularity is considered as a point
  existing outside of space-time and is thus removed from
  consideration.  However, this way of handling such entities is not
  appropriate, because then the Schwarzschild gravity described by Eq.
  (1$\cdot$1), for example, has to be regarded as produced by zero
  energy-momentum.\cite{Balasin1} A singularity exists in the
  space-time manifold, although it is not a \lq\lq place".\cite{Wald}
  We believe it constructive to examine space-time singularities by
  extending the analysis on manifolds in a distribution theoretical
  way, although we are not equipped with satisfactory mathematical
  machinery as yet. The expressions for scalar concomitants of the
  curvature tensor imply that mathematical formalism capable of
  incorporating objects like $\left[\delta^{(3)}(\mbox{\boldmath
      $x$)}\right]^2$ and $\delta^{(3)}(\mbox{\boldmath $x$})/r^3$ is
  needed for a precise description of singularities.
\item[{[I]}]\ We could not construct well-defined expressions for the
  quadratic scalars ${\hat{R}}^{\mu\nu}(\{\})$\linebreak
  $\times{\hat{R}}_{\mu\nu}(\{\})$ and
  ${\hat{R}}^{\rho\sigma\mu\nu}(\{\})
  {\hat{R}}_{\rho\sigma\mu\nu}(\{\})$.  Otherwise, however, our
  prescription leads to reasonable results, such as Eqs. (1$\cdot $3)
  and (3$\cdot $10). Thus, the claim, \lq\lq the well-known expression
  $\hat{R}^{\rho\sigma\mu\nu}(\{\})\hat{R}_{\rho\sigma\mu\nu}(\{\})
  =12a^2/\hat{r}^6$ is not correct $\cdots$'' in the statement (1.B),
  is well-grounded.
\item[{[J]}]\ Both {\bf Case 1} and {\bf Case 2} do not satisfy the
  harmonic condition, and {\bf Case 1} with $\Lambda =1=|\Omega |$ is
  suited for the description of the Schwarzschild gravity and its
  source. This is opposed to the argument in favor of the harmonic
  condition as a physical supplement to Einstein's theory of
  gravitation.\cite{Chou1,Peiyuan,Chou2,Nakanishi} For Lanczos
  coordinates, which satisfy the harmonic condition, the
  energy-momentum density of the source does not take the simple form
  Eq. (4$\cdot $10).  Harmonic coordinates are suited to discuss the
  gravitational wave, but they are not suited for the description of
  the Schwarzschild space-time.
\end{description}

\appendix

\section*{Appendix A \\
  {--- \small\emph{Proofs of relations (3$\cdot$8) and (4$\cdot$20)
      ---}}}
\addtocounter{section}{1}\setcounter{equation}{0}

First, we prove the relation (3$\cdot $8).  Let us define\footnote{In
  this Appendix, the symbol \lq\lq\^{}" is omitted for simplicity.}
\begin{equation}
F^{\alpha \beta }(\mbox{\boldmath $x$}; \epsilon )
\stackrel{\mbox{\scriptsize def}}{=}
\frac{{x}^\alpha {x}^\beta}{{r}^2}\left[%
\frac{3}{2}\frac{\epsilon^2}{({r}^2+\epsilon^2)^{5/2}}
+\frac{\epsilon^2}{{r}^2({r}^2+\epsilon^2)^{3/2}}\right]\;,
\end{equation}
and let $\Phi(\mbox{\boldmath $x$})$ be an arbitrary function of 
class $C^{\infty}$ with compact support: 
\begin{equation}
r^{2}\Phi^{\alpha \beta }(r)\stackrel{\mbox{\scriptsize def}}{=}
\int_{0}^{\pi }\sin \theta d\theta \int_{0}^{2\pi }
x^{\alpha }x^{\beta }\Phi(\mbox{\boldmath $x$})d\phi =0\;,
\; {\rm for}\; r\geq R 
\end{equation}
with $R$ being a positive constant.
Then, we have 
\begin{equation}
I^{\alpha \beta }(\epsilon )\stackrel{\mbox{\scriptsize def}}{=}
\int F^{\alpha \beta }(\mbox{\boldmath $x$}; \epsilon)
\Phi(\mbox{\boldmath $x$})d^{3}\mbox{\boldmath $x$}
=\int_{0}^{\frac{R}{\epsilon}}\Phi^{\alpha \beta }(\epsilon x)
\left[\frac{5}{2}\frac{1}{(x^{2}+1)^{3/2}}
-\frac{3}{2}\frac{1}{(x^{2}+1)^{5/2}}\right]dx\;,
\end{equation}
which is calculated to give
\begin{eqnarray}
I^{\alpha \beta }(\epsilon)&=&
\sum_{l=0}^{n-1}\frac{\Phi^{\alpha \beta (l)}(0)}{l!}\epsilon^{l}
\left[\frac{5}{4}B_{\frac{R^{2}}{R^{2}+\epsilon^{2}}}
\left(\frac{1+l}{2}, 1-\frac{l}{2}\right)-\frac{3}{4}B_{\frac{R^{2}}
{R^{2}+\epsilon^{2}}}\left(\frac{1+l}{2}, 2-\frac{l}{2}\right)\right]
\nonumber \\
&&+\frac{\epsilon^{n}}{n!}\int_{0}^{\frac{R}{\epsilon}}
\Phi^{\alpha \beta (n)}(\xi )x^{n}
\left[\frac{5}{2}\frac{1}{(x^{2}+1)^{3/2}}
-\frac{3}{2}\frac{1}{(x^{2}+1)^{5/2}}\right]dx \;, 
\end{eqnarray}
where we have expressed the function 
$\Phi^{\alpha \beta }(\epsilon x)$ as
\begin{eqnarray}
\Phi^{\alpha \beta }(\epsilon x)&=&\sum_{l=0}^{n-1}
\frac{\Phi^{\alpha \beta (l)}(0)}{l!}(\epsilon x)^{l}
+\frac{1}{n!}(\epsilon x)^{n}\Phi^{\alpha \beta (n)}(\xi )\;,
\nonumber \\  
& &\xi \stackrel{\mbox{\scriptsize def}}{=}\theta \epsilon x\;, 
\; \; 1>\theta >0\;, \; \; n\geq 1
\end{eqnarray}
with $\Phi^{\alpha \beta (l)}\stackrel{\mbox{\scriptsize def}}{=}
d^{l}\Phi^{\alpha \beta }/dr^{l}$. Also, $B_{z}(p,q)$ represents the 
incomplete beta function of the first kind:
\begin{equation}
B_{z}(p,q)\stackrel{\mbox{\scriptsize def}}{=}
\int_{0}^{z}t^{p-1}(1-t)^{q-1}dt\;, 
\; \; 1>{\rm Re}\, z>0\;.
\end{equation}
Equation (A$\cdot $4) gives
\begin{equation}
\lim_{\epsilon \rightarrow 0}I^{\alpha \beta }(\epsilon)
=2\pi \delta ^{\alpha \beta }\Phi (\mbox{\boldmath $x$}
=\mbox{\boldmath $0$})\;,\end{equation}
where use is made of the relation
\begin{equation}
\Phi^{\alpha \beta }(0)=\frac{4}{3}\pi \delta^{\alpha \beta }
\Phi (\mbox{\boldmath $x$}=\mbox{\boldmath $0$})\;.
\end{equation}
Thus, we see that $\lim_{\epsilon \rightarrow 0}F^{\alpha \beta}
(\mbox{\boldmath $x$}; \epsilon )=2\pi \delta^{\alpha \beta }
\delta^{(3)}(\mbox{\boldmath $x$})$. This is equivalent to 
Eq. (3$\cdot $8).

Next, we prove Eq. (4$\cdot$ 20). Let us define
\begin{equation}
G^{\alpha }(\mbox{\boldmath $x$};\epsilon)
\stackrel{\mbox{\scriptsize def}}{=}\frac{x^{\alpha }}{r}
\frac{\epsilon^{2}}{(r^{2}+\epsilon^{2})^{5/2}}\;,
\end{equation}
and let $\Psi(\mbox{\boldmath $x$})$ be an arbitrary function of 
class $C^{\infty}$ with compact support: 
\begin{equation}
r\Psi^{\alpha }(r)\stackrel{\mbox{\scriptsize def}}{=}
\int_{0}^{\pi }\sin \theta d\theta \int_{0}^{2\pi }
x^{\alpha }\Psi(\mbox{\boldmath $x$})d\phi =0\;,\; \;{\rm for}\; \; 
r\geq R>0\;. 
\end{equation}
Then, we have 
\begin{equation}
J^{\alpha }(\epsilon)\stackrel{\mbox{\scriptsize def}}{=}
\int G^{\alpha }(\mbox{\boldmath $x$}; \epsilon)
\Psi(\mbox{\boldmath $x$})d^{3}\mbox{\boldmath $x$}
=\int_{0}^{\frac{R}{\epsilon}}\Psi^{\alpha }(\epsilon x)
\frac{x^{2}}{(x^{2}+1)^{5/2}}dx\;,
\end{equation}
from which the expression 
\begin{eqnarray}
J^{\alpha }(\epsilon)&=&\frac{1}{2}
\sum_{l=0}^{n-1}\frac{\Psi^{\alpha (l)}(0)}{l!}\epsilon^{l}
B_{\frac{R^{2}}{R^{2}+\epsilon^{2}}}
\left(\frac{3+l}{2}, 1-\frac{l}{2}\right)
\nonumber \\
&&+\frac{\epsilon^{n}}{n!}\int_{0}^{\frac{R}{\epsilon}}
\Psi^{\alpha (n)}(\xi )\frac{x^{n+2}}{(x^{2}+1)^{5/2}}dx
\;, \; \; 
\theta R\geq \xi \geq 0\;,
\end{eqnarray}
follows, where we have expressed the function 
$\Psi^{\alpha }(\epsilon x)$ as
\begin{eqnarray}
\Psi^{\alpha }(\epsilon x)&=&\sum_{l=0}^{n-1}
\frac{\Psi^{\alpha (l)}(0)}{l!}(\epsilon x)^{l}
+\frac{1}{n!}(\epsilon x)^{n}\Psi^{\alpha (n)}(\xi )\;,\nonumber \\ 
& &\xi \stackrel{\mbox{\scriptsize def}}{=}\theta \epsilon x\;, 
\; \; 1>\theta >0\;, \; \; n\geq 1
\end{eqnarray}
with $\Psi^{\alpha (l)}\stackrel{\mbox{\scriptsize def}}{=}
d^{l}\Psi^{\alpha }/dr^{l}$. Equation (A$\cdot $12) leads to
$\lim_{\epsilon \rightarrow 0}J^{\alpha }(\epsilon)
=\Psi^{\alpha }(0)/3=0$, which gives Eq. (4$\cdot$20).

We note that the definite relation
\begin{equation}
\lim_{\epsilon \rightarrow 0}
\sum_{\alpha ,\, \beta =1}^{3}\frac{x^{\alpha }x^{\beta }}{r^{2}}
F^{\alpha \beta }(\mbox{\boldmath $x$}; \epsilon)
=6\pi \delta^{(3)}(\mbox{\boldmath $x$})\;, 
\end{equation}
exists, while, the limit
\begin{equation}
\sum_{\alpha ,\, \beta =1}^{3} \frac{x^{\alpha }x^{\beta }}{r^{2}}
\lim_{\epsilon \rightarrow 0}F^{\alpha \beta }(\mbox{\boldmath $x$};
\epsilon)\;, 
\end{equation}
is not well-defined, as is known from the footnote
\ref{footnote:ill-defined} on page \pageref{footnote:ill-defined}.
Also, we have
\begin{eqnarray}
\lim_{\epsilon \rightarrow 0}\sum_{\alpha =1}^{3} 
\frac{x^{\alpha }}{r}
G^{\alpha }(\mbox{\boldmath $x$};\epsilon)&=&\frac{4}{3}\pi
\delta^{(3)}(\mbox{\boldmath $x$})\neq 0
=\sum_{\alpha =1}^{3}\frac{x^{\alpha }}{r}
\lim_{\epsilon \rightarrow 0}
G^{\alpha }(\mbox{\boldmath $x$};\epsilon)\;.
\end{eqnarray}
We must be careful in treating 
$F^{\alpha \beta }(\mbox{\boldmath $x$};\epsilon)$ and 
$G^{\alpha }(\mbox{\boldmath $x$};\epsilon)$.

\section*{Appendix B \\
{--- \small\emph{Expressions for the quantities 
$\widetilde{\mbox{\boldmath $t$}}_\mu^{\ \nu}$, 
$\widetilde{\mbox{\boldmath $T$}}_\mu^{\ \nu}$, 
$R(\{\})$, $R^{\mu\nu}(\{\})R_{\mu\nu}(\{\})$ and \\
$R^{\rho\sigma\mu\nu}(\{\})R_{\rho\sigma\mu\nu}(\{\})$
in terms of $A,B,C$ and $D$ ---}}}
\addtocounter{section}{1}\setcounter{equation}{0}

The energy-momentum density 
$\widetilde{\mbox{\boldmath $t$}}_\mu^{\ \nu}$ is expressed as
\begin{eqnarray}
2\kappa \widetilde{\mbox{\boldmath $t$}}_0^{\ \mu}
&=&{\Bbb G}\delta_{0}^{\ \mu}\;,\; \; 
2\kappa \widetilde{\mbox{\boldmath $t$}}_{\alpha}^{\ 0}
=\frac{BD}{\sqrt{\Delta}}\left[\frac{2B'D'}{BD}+
\left(\frac{B'}{B}\right)^{2}
+\frac{1}{r}\frac{\Delta'}{\Delta}\right]\frac{x^{\alpha }}{r}\;,
\nonumber \\
2\kappa \widetilde{\mbox{\boldmath $t$}}_{\alpha}^{\ \beta}
&=&{\Bbb G}\delta_{\alpha }^{\ \beta }+\frac{AB}{\sqrt{\Delta }}
\Biggl\{\frac{1}{2r}\frac{AC+D^{2}}{AB}\frac{\Delta'}{\Delta }
\left(\delta_{\alpha }^{\ \beta }+\frac{x^{\alpha }x^{\beta }}
{r^{2}}\right)\nonumber \\
&&\qquad\qquad\qquad -\left.\left[2\frac{A'B'}{AB}
+\left(\frac{B'}{B}\right)^{2}\right]
\frac{x^{\alpha }x^{\beta }}{r^{2}}\right\}\;,
\end{eqnarray}
where ${\Bbb G}$ is given by
\begin{equation}
{\Bbb G}=\frac{AB}{\sqrt{\Delta}}\left[%
\frac{A'B'}{AB}
+\frac{1}{2}\left(\frac{B'}{B}\right)^2
-\frac{1}{r}\frac{AC+D^2}{AB}\frac{\Delta'}{\Delta }\right]\;,
\end{equation}
and $\Delta \stackrel{\mbox{\scriptsize def}}{=}A(B+C)+D^2$. 
Also, for
$\widetilde{\mbox{\boldmath $T$}}_\mu^{\ \nu}$, we have 
\begin{eqnarray}
\kappa \widetilde{\mbox{\boldmath $T$}}_0^{\ 0}
&=&\frac{AB}{\sqrt{\Delta}}\Biggl[
\frac{1}{r}\left(
\frac{A'}{A}+3\frac{B'}{B}-\frac{\Delta'}{\Delta }\right)
-\frac{1}{r^2}\frac{AC+D^2}{AB}+\frac{B''}{B}\nonumber \\
&&\qquad\qquad -\frac{1}{4}\left(\frac{B'}{B}\right)^2 
+\frac{1}{2}\frac{A'B'}{AB}
-\frac{1}{2}\frac{B'\Delta'}{B\Delta }\Biggr], \nonumber \\
\kappa\widetilde{\mbox{\boldmath $T$}}_0^{\ \alpha }&=&0, \; \; 
\kappa\widetilde{\mbox{\boldmath $T$}}_{\alpha}^{\ 0}=
\frac{x^\alpha}{r}\frac{BD}{\sqrt{\Delta}}\left[%
\frac{1}{r}\left(2\frac{B'}{B}
-\frac{\Delta' }{\Delta }\right)
+\frac{B''}{B}-\frac{1}{2}\left(\frac{B'}{B}\right)^2
-\frac{1}{2}\frac{B'\Delta'}{B\Delta }\right], \nonumber \\
\kappa \widetilde{\mbox{\boldmath $T$}}_{\alpha}^{\ \beta}
&=&\delta_\alpha^{\ \beta}\frac{AB}{\sqrt{\Delta}}\Biggl[%
\frac{1}{r}\left(%
\frac{A'}{A}+\frac{B'}{B}-\frac{1}{2}\frac{\Delta' }{\Delta }
\right)+\frac{1}{2}\frac{A''}{A}+\frac{1}{2}\frac{B''}{B}
-\frac{1}{4}\left(\frac{B'}{B}\right)^2 \nonumber \\
&&\qquad\qquad +\frac{1}{2}\frac{A'B'}{AB}
-\frac{1}{4}\left(\frac{A'}{A}+\frac{B'}{B}\right)
\frac{\Delta' }{\Delta }\Biggr] \nonumber\\
& &+\frac{x^{\alpha} x^{\beta}}{r^2}\frac{AB}{\sqrt{\Delta}}\Biggl[%
\frac{1}{2r}\frac{\Delta' }{\Delta }
-\frac{1}{r^2}\frac{AC+D^2}{AB}
-\frac{1}{2}\frac{A''}{A}-\frac{1}{2}\frac{B''}{B}\nonumber \\
&&\qquad\qquad\qquad +\frac{1}{2}\left(\frac{B'}{B}\right)^2
+\frac{1}{4}\left(\frac{A'}{A}+\frac{B'}{B}\right)
\frac{\Delta' }{\Delta }\Biggr]\;.
\end{eqnarray}
For $R(\{\}), R^{\mu\nu}(\{\})R_{\mu\nu}(\{\})$ and 
$R^{\rho\sigma\mu\nu}(\{\})R_{\rho\sigma\mu\nu}(\{\})$, we have 
\begin{eqnarray}
R(\{\})&=&\frac{A}{\Delta}\left[\frac{2}{r}\left(-2\frac{A'}{A}-
3\frac{B'}{B}+\frac{\Delta' }{\Delta }\right)
+\frac{2}{r^{2}}\frac{AC+D^{2}}{AB}
-\frac{A''}{A}-2\frac{B''}{B}\right.\nonumber \\
& &\qquad \left.+\frac{1}{2}\left(\frac{B'}{B}\right)^{2}
-2\frac{A'B'}{AB}
+\left(\frac{1}{2}\frac{A'}{A}+\frac{B'}{B}\right)
\frac{\Delta' }{\Delta }\right]\;, \nonumber \\
R^{\mu\nu}(\{\})R_{\mu\nu}(\{\})&=&\frac{A^{2}}{\Delta^{2}}
\left(\frac{1}{2}\frac{A''}{A}
-\frac{1}{4}\frac{A'\Delta' }{A\Delta } 
+\frac{1}{2}\frac{A'B'}{AB}+\frac{1}{r}\frac{A'}{A}\right)^{2}
\nonumber \\
& &+2\frac{A^{2}}{\Delta^{2}}\Biggl[\frac{1}{r}
\left(\frac{1}{2}\frac{\Delta'}{\Delta }-\frac{A'}{A}
-2\frac{B'}{B}\right)+\frac{1}{r^{2}}\frac{AC+D^{2}}{AB}
-\frac{1}{2}\frac{A'B'}{AB} \nonumber \\
&&\qquad\quad -\frac{1}{2}\frac{B''}{B}
+\frac{1}{4}\frac{B'\Delta'}{B\Delta }\Biggr]^{2}\nonumber \\
& &+\frac{A^{2}}{\Delta^{2}}\left[\frac{1}{2}\frac{A''}{A}
-\frac{1}{4}\frac{A'\Delta'}{A\Delta }
+\frac{1}{2}\frac{A'B'}{AB}
+\frac{B''}{B}-\frac{1}{2}\left(\frac{B'}{B}\right)^{2}
\right. \nonumber \\
& &\qquad\quad\left.-\frac{1}{2}\frac{B'\Delta'}{B\Delta }
+\frac{1}{r}\left(\frac{A'}{A}-\frac{\Delta' }{\Delta }
+2\frac{B'}{B}\right)\right]^{2}\;,\nonumber \\
R^{\rho\sigma\mu\nu}(\{\})R_{\rho\sigma\mu\nu}(\{\})
&=&\frac{A^{2}}{\Delta^{2}}\left(\frac{A''}{A}
-\frac{1}{2}\frac{A'\Delta' }{A\Delta }\right)^{2}
+2\frac{A^{2}}{\Delta^{2}}\left(\frac{1}{r}\frac{A'}{A}
+\frac{1}{2}\frac{A'B'}{AB}\right)^{2}\nonumber \\ 
& &+4\frac{A^{2}}{\Delta^{2}}\left[\frac{1}{r}\frac{B'}{B}
-\frac{1}{r^{2}}\frac{AC+D^{2}}{AB}
+\frac{1}{4}\left(\frac{B'}{B}\right)^{2}\right]^{2}\nonumber \\
& &+2\frac{A^{2}}{\Delta^{2}}\left[\frac{1}{r}\left(\frac{A'}{A}
+2\frac{B'}{B}-\frac{\Delta'}{\Delta }\right)
+\frac{1}{2}\frac{A'B'}{AB}+\frac{B''}{B}\right. \nonumber \\
& &\qquad\quad \left.-\frac{1}{2}\left(\frac{B'}{B}\right)^{2}
-\frac{1}{2}\frac{B'\Delta' }{B\Delta }\right]^{2}\;.
\end{eqnarray}
In the above, we have defined 
$A'\stackrel{\mbox{\scriptsize def}}{=}dA/dr, 
A''\stackrel{\mbox{\scriptsize def}}{=}d^{2}A/dr^{2}$, etc.

\end{document}